\documentclass[conference]{IEEEtran}
\IEEEoverridecommandlockouts

\usepackage{cite}
\usepackage{amsmath,amssymb,amsfonts}
\usepackage{algorithmic}
\usepackage{graphicx}
\usepackage{textcomp}
\usepackage{xcolor}
\def\BibTeX{{\rm B\kern-.05em{\sc i\kern-.025em b}\kern-.08em
    T\kern-.1667em\lower.7ex\hbox{E}\kern-.125emX}}
\begin{document}

\title{ITU-T Y.2325: NGN Evolution Towards Future}
\author{\IEEEauthorblockN{Rashmi Kamran\IEEEauthorrefmark{1}, Shwetha Kiran\IEEEauthorrefmark{1}, Pranav Jha\IEEEauthorrefmark{1}, Rashmi Yadav\IEEEauthorrefmark{2}, Abhay Karandikar\IEEEauthorrefmark{1}\IEEEauthorrefmark{3}, Prasanna Chaporkar\IEEEauthorrefmark{1}}
\IEEEauthorblockA{Department of Electrical Engineering,
Indian Institute of Technology Bombay, India\IEEEauthorrefmark{1}\\
\IEEEauthorblockA{Samsung Research Institute India, Bangalore, India\IEEEauthorrefmark{2}\\
\IEEEauthorblockA{Secretary to the Government of India,
Department of Science \& Technology, New Delhi, India\IEEEauthorrefmark{3}\\
Email: rashmi.kamran@iitb.ac.in\IEEEauthorrefmark{1},
shwethak@iitb.ac.in\IEEEauthorrefmark{1},
pranavjha@ee.iitb.ac.in\IEEEauthorrefmark{1},
rashmi.y@samsung.com\IEEEauthorrefmark{2},\\
karandi@ee.iitb.ac.in\IEEEauthorrefmark{1,3},
chaporkar@ee.iitb.ac.in\IEEEauthorrefmark{1}}}}}


\maketitle

\begin{abstract}
International Telecommunications Union (ITU) defined Next Generation Network (NGN) underlies most wireline and wireless packet-based telecommunications networks. A key design principle of NGN is decoupling of service-related functions from the underlying transport stratum, making user services independent of transport technologies. Interestingly, the NGN architecture, as defined in ITU standards, did not follow this design principle for internal network services, e.g., mobility, or authentication though adhering for external user services like IPTV or Multimedia services. These internal services are handled by the NGN transport control plane, making them an intrinsic part of the transport stratum, resulting in a tightly coupled service and transport functionality as opposed to the proclaimed design goal. This design choice may force each transport technology to support internal services individually, e.g., separate authentication service for each transport, leading to duplication. Since the NGN architecture is the base underlying architecture for most packet-based telecommunications network including advanced cellular networks like 4th/5th Generation cellular networks, the limitation persists in these cellular networks as well. To remedy the situation, the decoupling of service and transport can be generalized to include internal services like mobility and authentication also. In this context, the recently published ITU Y.2325 recommendation, defines an evolved NGN architecture, wherein all services, including internal network services, are decoupled from the transport stratum. The proposal results in a more scalable and modular evolved NGN architecture that can be used as a template for all future telecom networks including IMT-2030 (6th generation mobile networks). In this article, we review the evolved NGN architecture, as proposed in ITU-T Y.2325.
\end{abstract}


\section{Introduction}
\IEEEPARstart{I}{nternational} Telecommunication Union (ITU) Telecommunication Standardization Sector (ITU-T) defined Next Generation Network (NGN) is the underlying architecture for most broadband wireline and wireless packet-based telecommunications networks \cite{rec.y.2012}. 
As described in [ITU-T Recommendation Y.2001, 
``General overview of NGN"] \cite{rec.y.2001}, {\textit{NGN is a packet-based network able to provide Telecommunication Services to users and able to make use of multiple broadband, QoS-enabled transport technologies and in which service-related functions are independent of the underlying transport-related technologies. 
}}

\par \color{black}Although NGN architecture is capable of supporting diverse applications and services; rethinking its applicability to future networks is required due to significant changes in the telecom network landscape in recent years,
e.g., the emergence of IMT-2020 (Fifth  Generation (5G) wireless networks) and the  upcoming IMT-2030 and beyond (Sixth  Generation (6G)) wireless networks. Due to humongous increase in the number of users \cite{user_stats}, new use cases and increased user density (e.g., massive machine type communication), the signaling and data load on wireless networks has increased considerably. Signaling load on wireless networks is expected to increase further in the 6G era (e.g., ubiquitous connectivity). 
A need was felt by ITU-T to enhance the capabilities of the NGN by adopting technologies such as Software Defined Networking (SDN), Network Function Virtualization (NFV), distributed ledger technologies, and machine learning. Incorporation of these technologies is expected to pave the way for NGN evolution and allow it to better serve as the template for future networks. In this context, ITU-T initiated a study for NGN evolution in 2022 under its `Study Group 13 (SG13) - Future networks and emerging network technologies', wherein standards and recommendations with the objective to bring added flexibility, scalability, agility, programmability and other advanced characteristics to the NGN are being developed \cite{sg13}. The recommendation (standard) ITU-T Y.2325 - ``Architectural evolution for the next generation network control plane by applying software-defined networking technology", was developed under the same SG13. The recommendation highlights the requirements for evolved NGN transport control plane. Further, it also specifies an evolved architecture for NGN with enhanced scalability, flexibility and modularity.
\par Decoupling of service handling from the underlying transport stratum is one of the cornerstones of the NGN architecture, enabling a user service to utilize diverse transport technologies. Unfortunately, the NGN architecture, as defined in the original ITU-T standards, does not follow this architectural principle for internal (built-in) network services, e.g., mobility or authentication. These internal services are handled by the NGN transport (control) functions, making them an intrinsic part of the transport stratum, resulting in a tightly coupled service and transport functionality for the internal services. This design choice may force each transport technology to support internal services individually, e.g., separate authentication service for each transport leading to duplication of functionality. An example of such a duplication is found in 3rd Generation Partnership Project's (3GPP) 4th Generation (4G) mobile network, where devices using 3GPP access and non-3GPP access are authenticated separately. The design choice also results in a poor alignment with the SDN framework, where control plane is not really expected to directly handle user services of any type \cite{rec.y.3300}. The ITU-T Y.2325 recommendation improves the NGN architecture by pushing the handling of (internal) user services like mobility and authentication (along with the accompanying signaling exchange with devices) out of the transport control plane and making them part of the service stratum similar to user data, voice, or video services. This effectively results in treating signaling exchange with user devices as a user service (or data) in the evolved NGN (eNGN) architecture.

A few of the research works related to architecture-level evolutions for NGN are discussed next: An SDN-based service management model for NGN is proposed in \cite{ngn2}, which supports dynamic and automatic adjustment of Quality of Service (QoS) policies for better data transmission quality. A service-oriented architecture for NGN is proposed in \cite{ngn3} specifically rebuilt to ensure the massive scalability of multimedia services and to fulfill their stringent requirements. In \cite{ngn4}, a three-tier NGN architecture is proposed to address mobility-related challenges in NGN and also provides details of handover functionalities at the access level. A cloudified NGN infrastructure based on SDN and Network Function Virtualization (NFV) is discussed in order to improve network scalability, elasticity, usage efficiency, and resiliency while decreasing cost \cite{ngn5}. However, it does not provide any performance evaluation to verify the advantages. A QoS-Aware resource and admission control and management method for NGN is investigated in \cite{ngn6} to improve the scalability of transport networks. Nominal research work focuses on the evolution of NGN architecture. The above-discussed works are not very recent, and to the best of our knowledge, no prior art is available for NGN evolution towards handling usage scenarios and requirements defined for IMT 2030. 

In this paper, we present learnings from the ITU-T recommendation, Y.2325. We also present the detailed working flow as specified in Y.2325. Further to demonstrate the advantages of the evolved NGN architecture as presented in Y.2325, we present scalability performance of the same viz-a-viz the existing NGN architecture.

Rest of the paper is organized as follows. Section\,\ref{idea} provides an overview of the evolved NGN control plane. It also provides a general comparison for handling of built-in services in existing and evolved NGN.
Sections \ref{highn} and \ref{highm} describes high level functions for network attachment and mobility services respectively. Section \ref{flows} provides an example information flow for mobility service in evolved NGN. Scalability performance evaluation is presented in Section \ref{results}. Section \ref{benefits} presents some benefits of the proposed architecture and section \ref{conclusion} concludes the findings based on learnings and evaluation.

\section{Evolved NGN control plane}\label{idea}
\begin{figure}[!h]
\centering
\vspace{-0.5cm}
\includegraphics[width=3.5in]{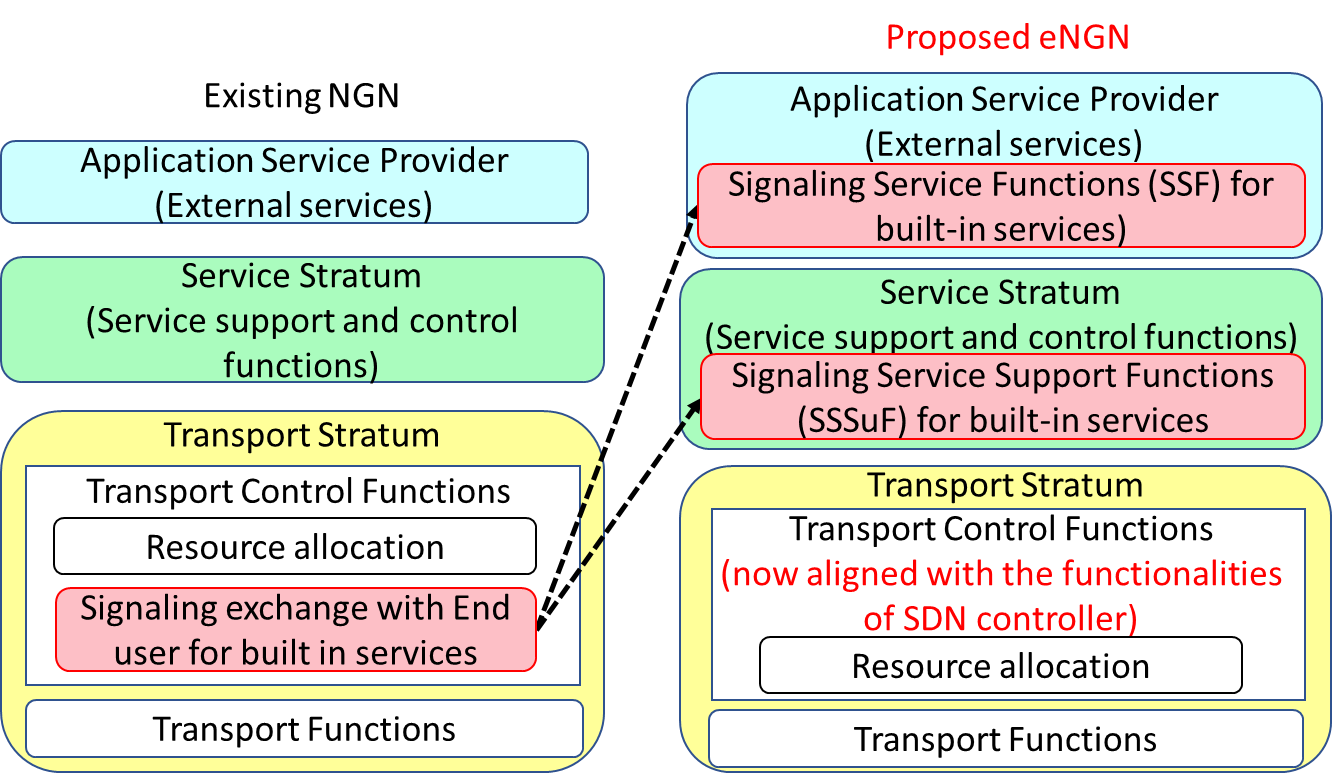}
\vspace{-0.4cm}
\caption{Conceptual difference between existing NGN and evolved NGN. Treating signaling exchange with end user as a service in the evolved NGN.}
\vspace{-0.4cm}
\label{ngne}
\end{figure}

The main conceptual difference between existing NGN and evolved NGN (as specified in Y.2325) is shown in Fig. \ref{ngne}. The functionalities of user plane control (resource allocation) and signaling exchange with the end user are tightly coupled in the existing NGN transport control plane. In the evolved NGN control plane, these two functionalities are decoupled, and the end user signaling is handled using separate Signaling Service Functions (SSFs) and Signaling Service Support Functions (SSSuFs) for built-in services, as shown in the right-hand side of Fig.\ref{ngne}. In addition, end user signaling can be treated as another form of data in the network. As a result, controlling and managing transport functions is the primary responsibility of the transport control functions in the evolved NGN, which involves user session management (i.e., establishment, modification, or deletion) via the transport functions (user plane). This subdivision of functionality is named `resource allocation (control) functionality' in Fig. \ref{ngne}. However, the functionality of signaling exchange with the end user is separated and moved out of the transport control plane and is considered a built-in service in the application stratum. 
\par Considering the baseline described above, the ITU-T Y.2325 recommendation introduces new SSFs to facilitate various built-in services.  The built-in services (using associated SSFs and SSSuFs) are handled in the same way as external applications are handled in the network.  In addition, the evolved NGN transport control plane retains the resource allocation control functionality, which is now fully aligned with the functionalities of the SDN control plane (SDN controller).  Communication (signaling) between end users and SSFs occurs over an established data path (session).  Transport control functions shall interact with the SSFs and not directly with the end users to provide built-in services to the end users.  SSFs are now responsible for exchanging information with end users through a session.  These functions serve built-in services directly by communicating with the end user or indirectly by managing those services.  For example,  in the case of a network-based mobility scenario, mobility-associated SSF receives location/connectivity-related updates from the end user and takes a decision related to handover. Further, SSF requests resource control functions to allocate resources accordingly.  Signaling service support functions are also introduced in the service stratum to support signaling service functions of built-in services.  Simple examples of built-in services in NGN are network attachment and mobility management. 
\subsection{Handling of built-in services in NGN viz-a-viz eNGN} 
\begin{figure}[h!]
\centering
\vspace{-0.2cm}
\includegraphics[width=3.5in]{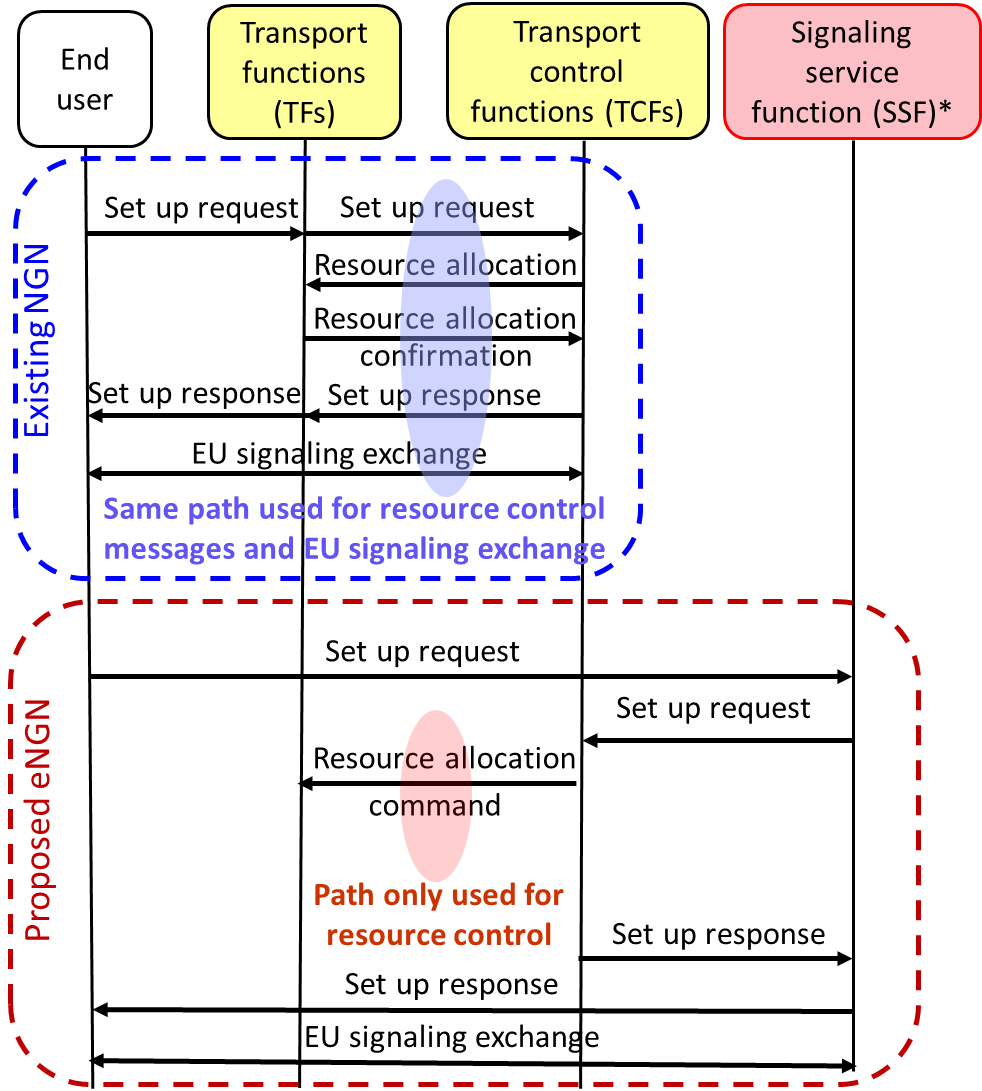}
\vspace{-0.5cm}
\caption{Handling of built-in services in the existing NGN and proposed eNGN. *Please note that Signaling service functions (SSFs) are orchestrated through Signaling Service Support Functions (SSSuFs)}
\vspace{-0.2cm}
\label{handling}
\end{figure}

The existing NGN does not handle built-in and external services in a uniform way. Whereas, the evolved NGN proposal results in uniform handling of all types of services (including built-in and external services). Figure \ref{handling} shows the high-level flow for end user signaling exchange for built-in services for both existing NGN and eNGN. In the existing NGN, when the end user sends a setup request for any built-in service, transport functions send resource allocation control messages to transport functions. Upon receiving resource allocation confirmation from transport functions, transport control functions send a setup response to the end user. And end user signaling exchange related to built-in service occurs between transport control functions and the end user. Please note that the same protocol/path between transport control and transport functions is used for resource control messages and end user signaling exchange.
\par A new SSF (as an application/service in application stratum) specific to each built-in service  along with a signaling service support function (in service stratum) has been introduced in the eNGN to serve that built-in service. In eNGN, the end user sends a setup request to the SSF for a built-in service. The SSF requests transport control functions for resource allocation for sessions between the end user and SSF. Transport control functions are responsible for resource allocation control and send setup response to confirm resource allocation after sending resource allocation commands to transport functions. Now, a session is available between SSF and the end user for signaling exchange. Through this session,  SSF sends a setup response, and this session can be used for end user signaling exchange required for built-in service. The main difference between NGN and eNGN is that the interface between transport control functions and transport functions is more scalable in the case of eNGN as it is only responsible for resource allocation commands. 
\subsection{Functional Decomposition and Placement in eNGN}
\begin{figure}[!h]
\centering
\vspace{-0.2cm}
\includegraphics[width=3.2in]{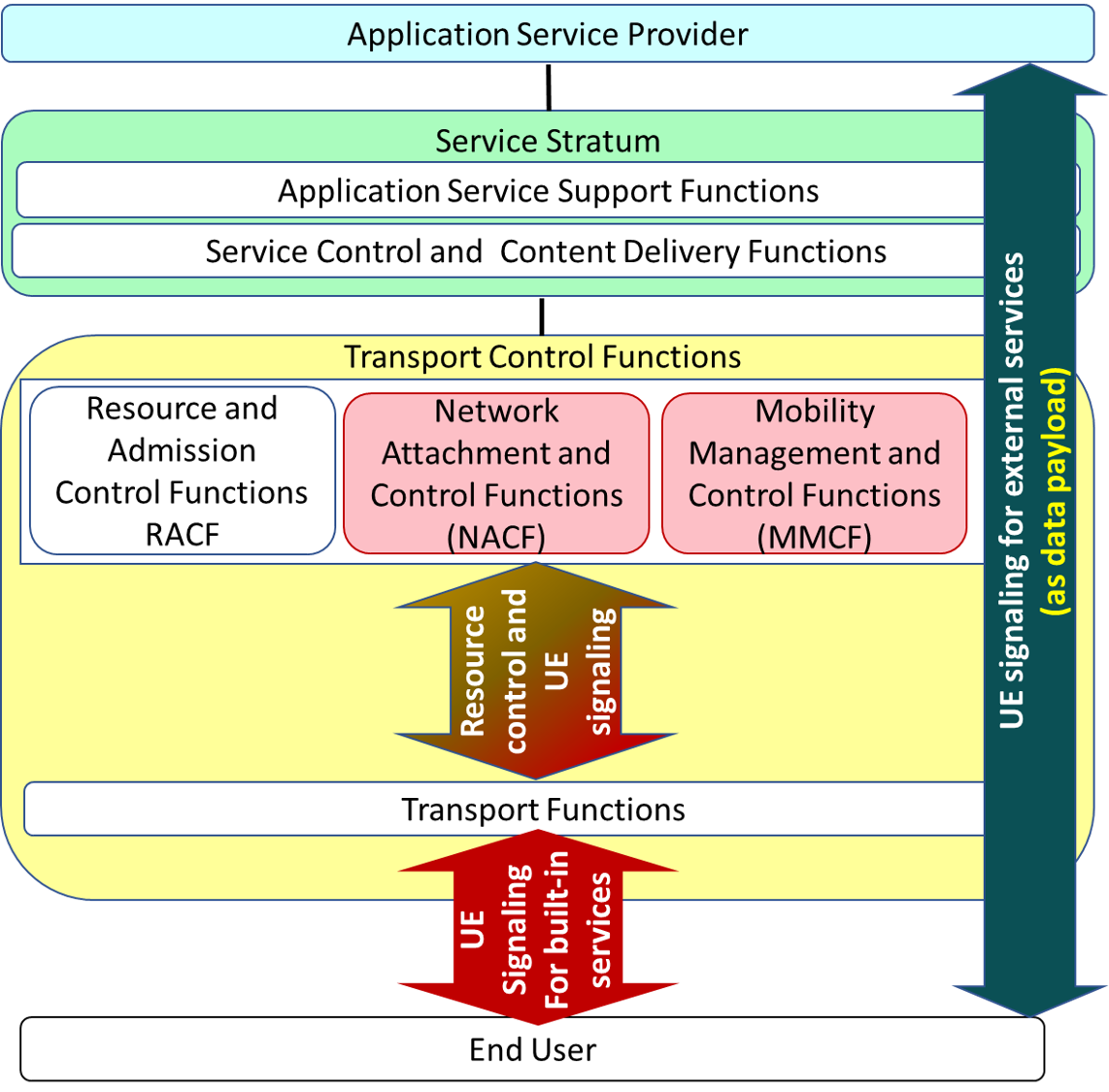}
\vspace{-0.2cm}
\caption{Signaling exchange through transport control functions in NGN}
\vspace{-0.2cm}
\label{existing}
\end{figure}

\begin{figure}[!h]
\centering
\vspace{-0.2cm}
\includegraphics[width=3.2in]{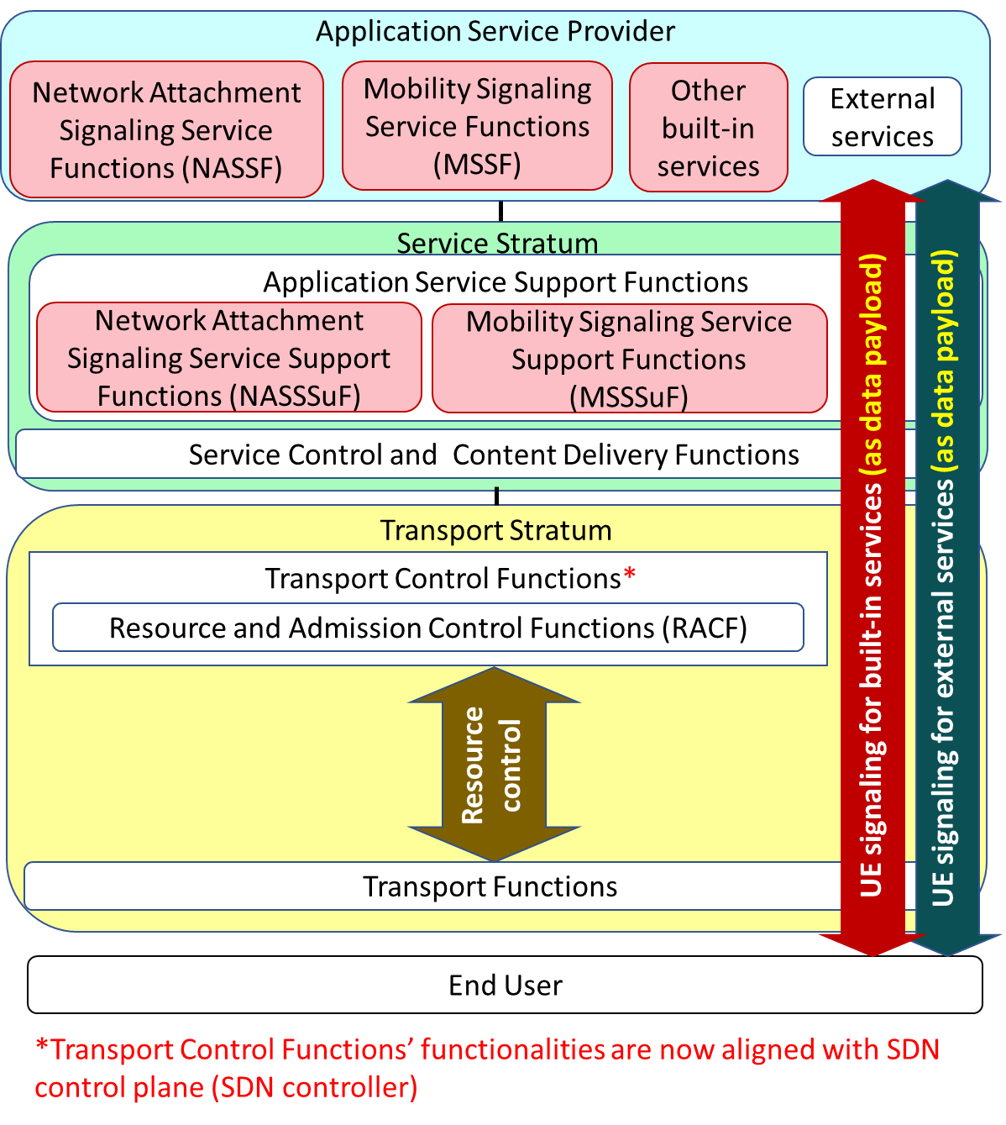}
\vspace{-0.2cm}
\caption{Placement of High Level Functions in the Evolved NGN Architecture and signaling exchange with end user as data payload}
\vspace{-0.3cm}
\label{proposed}
\end{figure}

Figures \ref{existing} and \ref{proposed} provide an overview of how the existing NGN control plane architecture transformed into the eNGN control plane. In the existing NGN architecture (as shown in Fig. \ref{existing}), the control plane (or signaling) functionalities for in-built services like mobility management and network attachment control functions are placed in the transport stratum (transport control functions). These functions interact with the end users to collect information and execute mobility and network attachment-related procedures. In contrast, external applications communicate with the control plane via the service stratum's functions (i.e.,  Application Support Functions (ASF) and Service Support Functions (SSF)) and transport control functions (which is part of the transport stratum) to facilitate signaling information exchange with the end users. This complex intertwined structure is simplified by applying SDN technology to develop the eNGN control plane architecture.The redesigned transport control functions in eNGN correspond to an SDN controller. Signaling service functions that are placed in the application stratum collect information/requirements from the end user. Two steps are involved in the completion of a service: first is the collection of service requirements from the end user, and second is the establishment of the data path (session) for service delivery. Nevertheless, a data path (session) needs to be established to collect the service requirements from the end user as well.

\par Transport control functions in existing NGN architecture consisted of the following high-level functions: Resource and Admission Control Functions (RACF), Network Attachment Control Functions (NACF), and Mobility Management and Control Functions (MMCF). In evolved NGN architecture, built-in services like network attachment and mobility management services are handled similarly to external applications/services. Hence, these two functions are split into SSFs and SSSuFs and moved out of the transport control. The SSF for network attachment service is called the Network Attachment Signaling Service Function (NASSF), and for the mobility management service is called the Mobility Signaling Service Function (MSSF); they are placed in the application stratum. Similarly, the SSSuF for network attachment service is called the Network Attachment Signaling Service Support Function (NASSSuF), and for the mobility management service is called Mobility Signaling Service Support Function (MSSSuF); these functions provide support for NASSF and MSSF, respectively and are placed in the service stratum. Figure \ref{proposed} provides a detailed overview of how these functions are positioned in the evolved NGN architecture.

\section{Network Attachment in evolved NGN}\label{highn}
Detailed functionalities of the NACF for the existing NGN and its functional entities are defined in ITU-T Y.2014 \cite{rec.y.2014}. The NACF supports end user registration at the access level and initializes the end user to allow access to the NGN services. The NACF also provides network-level authentication, authenticating the access sessions and managing the access network's IP address space. NACF also informs the end users about the contact points in the NGN service stratum. All these functionalities are supported in the evolved NGN architecture via the NASSF and NASSSuF. 
\subsection{Network Attachment Signaling Service Function (NASSF)}
NASSF, being a built-in service, communicates with the end user via an established data session. NASSSuF (in service stratum) supports NASSF to establish this data path / session before the network attachment services are provided. Detailed functionalities of NASSF are as follows:
\begin{itemize}
    \item{It supports end user registration at the access level and initializes end users to access NGN services.}
    \item{It is responsible for IP Address allocation to the end user, including permanent and temporary IP addresses. A temporary IP address is needed to support the mobility of an end user.}
    \item{Authentication and authorization support is also a responsibility of NASSF. It also holds end user subscription authentication data such as identifiers.}
    \item{It handles the association of the end user location with IP address information. This information is sent to other network function queries.}
    \item{It provides network configuration parameters such as Domain Name System (DNS) info to end users.}
    \item{It is also responsible for initialization and configuration of Home Gateway (HGW), a gateway between the customer premises network and the access network. This feature is optional.}
    \item{It coordinates with NASSSuF to initialize a session with the end user equipment. This session allows network attachment services to be provided to the end users.}
\end{itemize}
\subsection{Network Attachment Signaling Service Support Function (NASSSuF)}
NASSSuF, in coordination with RACF (a transport control function), facilitates admission control and resource allocation in the data plane (transport functions) for NASSF. This function is a service support function for the  NASSF (Application stratum) that provides network attachment service to end users. The functionalities of NASSSuF are as follows:
\begin{itemize}
    \item{It supports session establishment between the NASSF and the end user by requesting RACF to configure resources in the transport plane.}
    \item{It collects end user associated access network information and provides it to NASSF.}
    \item{It can also serve to collect accounting data for each end user, though this feature is optional.}
\end{itemize}

\section{Mobility Management in evolved NGN}\label{highm}
Details of the MMCF functionalities in the existing NGN, along with its functional entities, are defined in ITU-T Y.2018 \cite{rec.y.2018}. MMCF supports two main functionalities: End User location management and handover control. Handover Control Functions (HCF) provide session continuity for ongoing sessions of an end user when it is moving. Both these functionalities are supported in the evolved NGN through MSSF and MSSsupportF. Please note that the eNGN supports two types of mobility scenarios (same as supported in existing NGN \cite{rec.y.2018}); `host-based mobility,' where the end user takes the lead on initiating mobility service (handover), and `network-based mobility,' where handover decision is taken by the network.
\subsection{Mobility Signaling Service Function (MSSF)}
MMSF is an SSF for mobility service (built-in service) that communicates with the end user via an established data session. MSSSuF (placed in the service stratum) helps MSSF establish this data path/session to support mobility services. MSSF also interacts with ASF and SSF in the service stratum. The functionalities of MSSF are categorized and detailed as follows:
\begin{itemize}
    \item{Location management: Location registration is initiated by the MSSF on behalf of the end user for network-based mobility. In the case of host-based mobility, the MSSF binds the persistent IP address with the temporary IP address of the end user. The MSSF also supports new location binding related aspects and provides this information to MSSSuF.}
    \item{Handover signaling: In network-based mobility, the MSSF sends the requests from the end user and forwards them to MSSSuF for further processing of handover. In host-based mobility, the MSSF sends the potential links that are acceptable for end user connectivity. The end user then triggers the handover procedure. }
    \item{Mobility service authentication and authorization: MSSF supports authentication and authorization of the end user for mobility services in coordination with NASSF. The end user indicates mobility service-related requests during the network attachment procedure. The end user also indicates host-based or network-based mobility preferences while registering for mobility services.}
    \item{Session initiation: MSSF via MSSSuF requests for the signaling session to be initiated with the end user to support mobility-related procedures.}
\end{itemize}
\subsection{Mobility Signaling Service Support Function (MSSSuF)}
MSSSuF is a service support function in the service stratum; it works with the transport control functions, i.e., RACFs in the data plane (transport functions), to provide mobility-related information and further for associated resource configuration. The functionalities of MSSSuF are detailed as follows:
\begin{itemize}
    \item{Handover Decision: In the case of network-based mobility, MSSSuF decides on handover aspects for the end user. With host-based mobility, the end user takes the decision on handover.}
    \item{Resource Configuration: MSSSuF works with RACF for resource configuration based on the handover decision. It ensures that the potential access links for the end user to handover can meet quality of service requirements for the data sessions. It also requests releasing old data path resources in the data plane (transport functions).}
    \item{Handover Information Sharing: MSSSuF forwards the handover policy information to the end user to support host-based mobility. This policy contains operator-defined preferences and rules that affect the handling of the handover.}
    \item{Provisioning of Network Information Repository: MSSSuF is responsible for maintaining neighboring networks information, which helps the end users with discovery and connectivity to access networks.}
\end{itemize}
\subsection{Coordination between signaling service functions of built-in services}\label{flows}
\begin{figure}[h!]
\centering
\vspace{-0.4cm}
\includegraphics[width=3in]{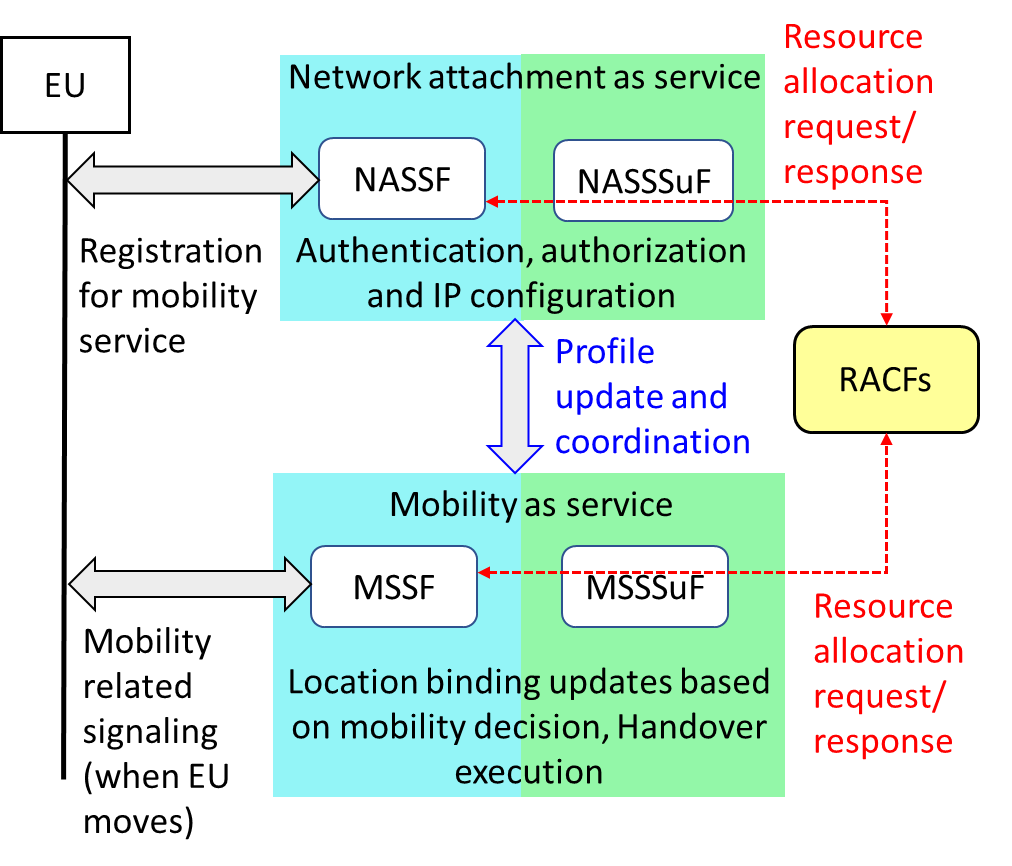}
\vspace{-0.4cm}
\caption{Coordination between signaling service functions of built-in services in eNGN}
\label{flow}
\vspace{-0.4cm}
\end{figure}

Figure \ref{flow} demonstrates the coordination between signaling service functions of built-in services by considering Network attachment and Mobility as service.  The end user registers for mobility service while registering to the network, through NASSF and NASSSuF. Authentication and authorization is performed by NASSF with the support of NASSSuF. Upon receiving request  from the end user for session establishment, NASSF handles IP configuration allocation and mapping of IP configuration with the logical identifier. Accordingly, resource allocation request for session establishment is sent to NASSSuF; it forwards this request to RACFs. RACF allocates resources for the session and sends a response to NASSSuF.  
When End user moves, it sends mobility location binding update requests to MSSF. MSSF performs location binding updates for the end user for handover completion and takes care of handover execution. 
NASSSuF also need to coordinate with MSSF for end user-related identities and for updating user profile. Main key takeaway is that all signaling exchanges with the end user related to network attachment and mobility are handled by signaling service functions (NASSF and MSSF). To facilitate this signaling exchange, a session is established between the end user and the signaling service functions (NASSF and MSSF) through signaling service support functions.

\section{Performance Evaluation: Scalability}\label{results} 
\begin{figure}[!h]
\centering
\vspace{-0.4cm}
\includegraphics[width=3.6in]{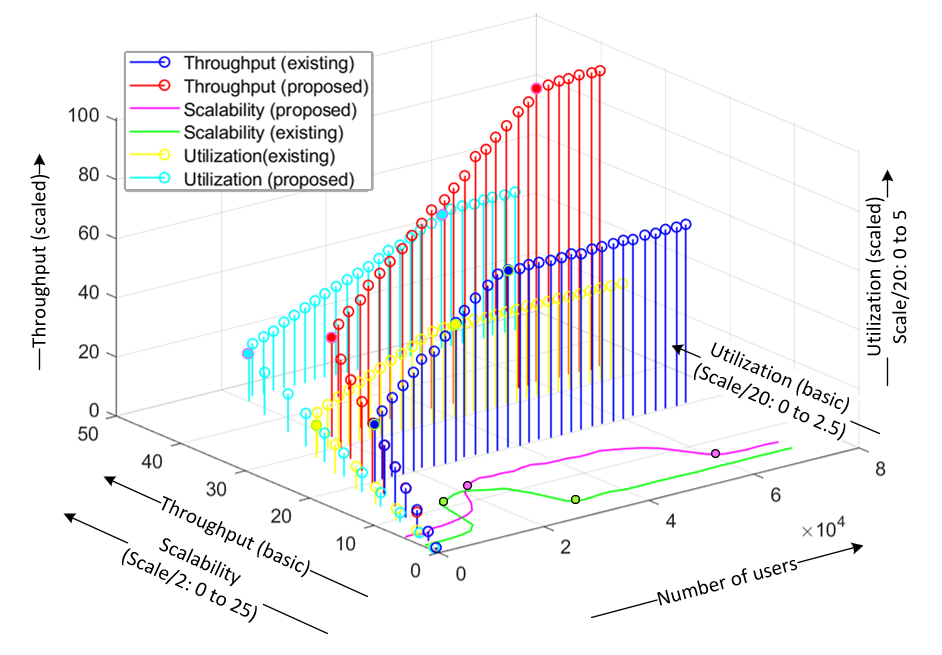}
\vspace{-0.9cm}
\caption{Scalability comparison between NGN and evolved NGN}
\label{scangne}
\end{figure}

Scalability metric (defined in \cite{Jogalekar}) is considered to compare the performance of existing NGN and eNGN. The scalability analysis involves various parameters such as throughput, processor utilization, and average response time. This metric gives a view of network load handling capacity with a scaled number of components. Figure \ref{scangne} shows the scalability performance comparison between the existing and evolved NGN. The high-level flow of handling built-in services in existing NGN and eNGN (Figure \ref{handling}) is modeled using PEPA software with basic (say, number of processors is N) and scaled (number of processors is 2N) configurations to evaluate scalability. PEPA is a process algebra-based language to model processors and their processing time and interactions for/between various components of any distributed system. For existing NGN, three components, the end user, transport functions, and transport control functions, are considered with corresponding processors. For eNGN, one more component, SSF, is considered. In simulations, the total number of processors used in the case of both existing and even architectures are kept equal for comparing performance with the exact hardware requirements. The existing NGN saturates at 14,000 and 38,000 users for basic and scaled configurations, and the eNGN saturates at 22,000 and 62,000 users for basic and scaled configurations, respectively. Saturation points observed in the throughput and utilization plots are aligned as saturation comes due to any of the processor's full utilization points. It can be observed from the scalability results that more users can be served by eNGN than by existing NGN for both configurations. The scalability factor is also improved in the case of eNGN compared to existing NGN. 
\section{Benefits of the evolved NGN control plane}\label{benefits} In addition to the improved scalability performance of eNGN over exiting NGN (as discussed in the above section), there are the following benefits of eNGN: 
\subsubsection{Enhanced modularity} Decoupling signaling handling functionality from the transport control functions results in enhanced modularity in eNGN. As dedicated signaling service functions handle the end user signaling, the signaling messages are carried by user plane (transport) functions in the path, i.e., the signaling service functions and the end users can exchange signaling messages via IP connectivity established through the transport functions in the network, which simplifies the signaling information flow in the network considerably.
\subsubsection{Transport agnostic built-in service handling} By decoupling the built-in service functions from the transport stratum, it is possible to have unified and transport agnostic handling of built-in services like mobility and authentication. This also avoids duplication of the built-in service  functionality.

\subsubsection{Simplified resource control protocols} In the existing NGN, transport control functions exchange signaling messages with end users. The path for this end user – transport control function communication passes through the transport functions. The transport control plane functions also need to control user plane functions using a resource control interface. For both types of communication (end user – transport control plane communication and transport functions – transport control plane communication), the transport control functions use the resource control protocols in the exiting NGN. It increases the complexity of the resource control protocols. In eNGN,  resource control protocols are simplified as they only need to support resource control messages.
\subsubsection{Slice/service specific signaling through dynamic deployment of signaling service functions} In eNGN, different signaling overheads for different slices/services are supported with ease through the deployment of slice-specific signaling service functions once the signaling handling functionality is separated from the transport control functions and independently deployed as signaling service functions in the NGN service stratum. 
\subsubsection{Uniform service delivery mechanism}  The eNGN provides uniform treatment to built-in and external services to simplify the procedures and architecture, which is not supported in the existing NGN. 
\section{Conclusion}\label{conclusion}
We review the ITU-T Y.2325 recommendation, specifying a scalable and simplified evolved NGN (eNGN) control plane architecture in which signaling exchange with the end user is considered a service. It leads to uniform handling of internal and external services and a better alignment of the network architecture with SDN technology. It provides use case-specific signaling support with the dynamic deployment of signaling service functions. We also verified the advantage of eNGN by comparing its scalability with that of existing NGN. This framework has the potential to be a baseline for designing IMT 2030 architectures to cater to diverse use case scenarios. Further, other built-in services like authentication and use case-specific flows can be evaluated using the proposed eNGN.

\bibliographystyle{IEEEtran}
\bibliography{NGNe.bib}

\end{document}